# Simulation Organogenesis in COMSOL: Deforming and Interacting Domains

Denis Menshykau[1], Dagmar Iber[*,1,2]
[1]D-BSSE, ETH Zurich, Switzerland; [2]SIB, Switzerland
*Corresponding author: D-BSSE, ETH Zurich, Mattenstrasse 26, 4058, Switzerland, dagmar.iber@bsse.ethz.ch

**Abstract:** Organogenesis is a tightly regulated process that has been studied experimentally for decades. We are developing mechanistic models for the morphogenesis of limbs, lungs, and kidneys with a view to integrate available knowledge and to better understand the underlying regulatory logic. In our previous paper on simulating organogenesis in COMSOL (German et al COMSOL Conf Proceedings 2011) we discussed methods to efficiently solve such models, predominantly on a static domain. Organ size, however, changes dramatically during development, and tissues are composed of several layers that may expand both together or independently. Moreover, the developing organ are typically embedded in an environment, and diffusional exchange with this environment can affect the patterning process. Here we discuss methods to build and efficiently solve models involving large deformation of composite domains in COMSOL Multiphysics.

## 1. Introduction: Mechanistic Models for Organogenesis

Organogenesis is a tightly regulated process. Many important genes have been identified and the regulatory logic has been analyzed extensively over the last decades. The discovered regulatory networks are too complex to be understood intuitively and many questions remain open. Computational models can help to integrate available knowledge, to test the consistency of current models, and to generate new hypotheses.

During organogenesis tissue layers organize and differentiate into functionally organized units. Recent studies demonstrate the highly deterministic nature of the developmental processes. Thus lung branching morphogenesis is extremely stereotyped such that the lung tree with its thousands of branches is identical in embryos in the same genetic background.[1] The proximal part of the bronchial tree is built by three geometrically simple, recursive modes of branching.[1] This suggests a deterministic underlying process, and deterministic models for pattern formation as studied for decades in developmental biology thus appear suitable.[2] We have previously developed and solved such models in COMSOL Multiphysics and we correctly predicted novel genetic regulatory interactions in the limb bud[3], and we suggested a mechanism for branch point selection, smooth muscle and vasculature morphogenesis in the lung[4,5] as well as for patterning during long bone development[6] (Figure 1). COMSOL has previously been shown to solve similar problems with a known analytic solution accurately.[7]

Our models are formulated as systems of reaction-diffusion equations of the form:

$$\dot{X}_i + \nabla(u \cdot X_i) = D_i \nabla^2 X_i + R_i$$

where $u$ denotes the velocity field of the domain and $R_i$ the reactions, which couple the equations for the different species $X_i$. $D_i$ is the diffusion constant and $\nabla$ the Nabla operator. The velocity field might be either imposed or be based on concentrations of proteins, which changes the behavior of the cells, e.g. the division rate or cell adhesion and motility. A wide range of reaction laws can be used[8]; typical reactions include

$$R_X = -\delta \cdot X$$

for the decay of component X and

$$R_X = -k^+ \cdot m \cdot X^m \cdot Y^n + k^- \cdot m \cdot X_m Y_n$$

$$R_Y = -k^+ \cdot n \cdot X^m \cdot Y^n + k^- \cdot n \cdot X_m Y_n$$

$$R_{X_m Y_n} = k^+ \cdot X^m \cdot Y^n - k^- \cdot X_m Y_n$$

for the formation of a complex $X_m Y_n$ made of $m$ X and $n$ Y molecules. The reaction terms can contain also other non-linear functions like enzymatic activation $\sigma$ and inhibition $\bar{\sigma} = 1 - \sigma$, where $\sigma$ is modelled analogous to Hill kinetics (Michaelis-Menten for n=1):

$$\sigma = X^n / (X^n + K^n).$$

The threshold *K* is the concentration at which the activation reaches half its strength and the exponent *n* depends on the cooperativity of the regulating interactions. For example

$R_X = \rho \cdot \sigma(Y)$

describes a production term for a protein *X* that is induced by another protein *Y*.

Our models typically comprise three to fifteen variables, and the discretized problem typically has 50 000 to 500 000 degrees of freedom.

## 2. Conventions & Computational Details

Here we use **bold italic** to refer to COMSOL fields and nodes e.g. *Coefficient form PDE*, *Diffusion Coefficient* refers to the field where diffusion coefficient need to be specified.

Models discussed in this manuscript are implemented in COMSOL 4.2a through the *Coefficient form PDE* interface; models featuring deforming domains are implement by coupling the *Coefficient form PDE* and *ALE Moving Mesh* interfaces.

The following COMSOL settings and options are used in the models discussed in this manuscript:
- *first order Lagrange shape function* for problem discretization;
- The *direct* solver *Pardiso*, variables are *segregated* at least into two groups: first group - variables defined in the PDE; second group - variables describing domain deformation (*.xyz* or *.xy*);
- Jacobian is set to be updated at every time step;
- to model implicitly defined domain deformation *Prescribed n velocity* was defined, while *Prescribed t1 velocity* was left undefined;
- to model implicitly defined deformation of a composite domain *Lagrange multipliers* were enabled at subdomain boundaries;

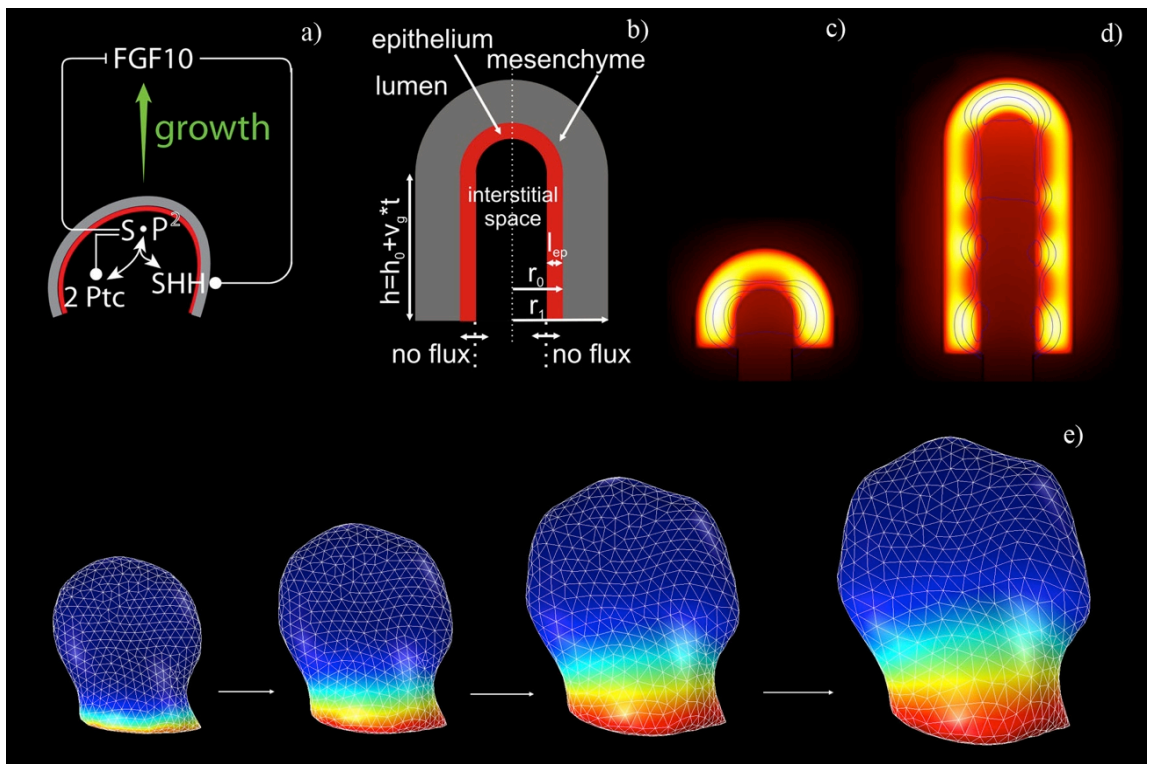

**Figure 1. Computational Models for Organogenesis in COMSOL**. The upper panel depicts a computational model for early lung branching morphogenesis. a) The core regulatory network for lung branching morphogenesis; b) the idealized computational domain comprises a 2D cross-section along the cylinder axis of symmetry; c) and d) computed distribution of FGF10 (color code) and SHH (contour lines) in bifurcation and lateral branching modes of branching. Brighter color depicts regions with higher concentration of FGF10. The lower panel shows the solution of a reaction-diffusion equation on a three dimensional deforming domain in the shape of embryonic limb bud as extracted from microscopic OPT data (courtesy of Erkan Uenal, Zeller lab, DBM, University of Basel).

- if *Lagrange multipliers* was enabled at the boundaries of deforming domain *Exclude algebraic* option was chosen in the *Advanced* section of time dependent solver;
- to model implicitly defined deformation the *Automatic Remeshing* solver feature was enabled.

## 3. Use of COMSOL Multiphysics
### 3.1 Static Composite Domains

Organs usually comprise several tissues e.g. embryonic lungs and kidneys are built of distinct layers of epithelial and mesenchymal cells. Both lung and kidney branching are controlled by signaling from the epithelium to the mesenchyme and back. From the computational (mathematical) point of view this means that different sets of PDEs describe the epithelial and mesenchymal layers. COMSOL provides several possibilities to define such a model: 1) a global set of PDEs is defined on the entire composite domain; 2) PDEs are defined on the subdomain and then coupled. PDEs defined on subdomains are coupled by imposing the following *Constraints*: $c^i_j$-$c^i_k$ on the appropriate subdomain borders, where the superscript index refers to a variable and the subscript index refers to a domain index. Figure 2a depicts a composbbite one-dimensional domain, where the diffusion coefficient differs on the subdomains.

In the first instance we define the PDE globally on a composite one-dimensional domain according to

$$\dot{L} = D_i \Delta L - \sigma L \quad (1)$$

with the boundary conditions
$L(0)=1$, $L_x(l_0)=0$
where, $D_i = 1$ on domain 1 and 100 on domain 2, $\sigma=1$.

This results in artifacts near the subdomain border: the concentration $L$ shows an artificial minimum (Figure 2b, red line) and the diffusional flux has a discontinuity (Figure 2c, red line).

In the second approach the PDE is defined locally on the subdomains:

$$\dot{L}_1 = D_1 \Delta L_1 - \sigma L_1$$
$$\dot{L}_2 = D_2 \Delta L_2 - \sigma L_2 \quad (2)$$

with boundary conditions
$L_1(0)=1$, $L_2(l_0)=0$ and $L_1(l_{border})=L_2(l_{border})$
where, $D_1 = 1$ and $D_2 = 100$ on domain 2, $\sigma=1$. This approach does not suffer from such artifacts (Figure 2b,c, black curves).

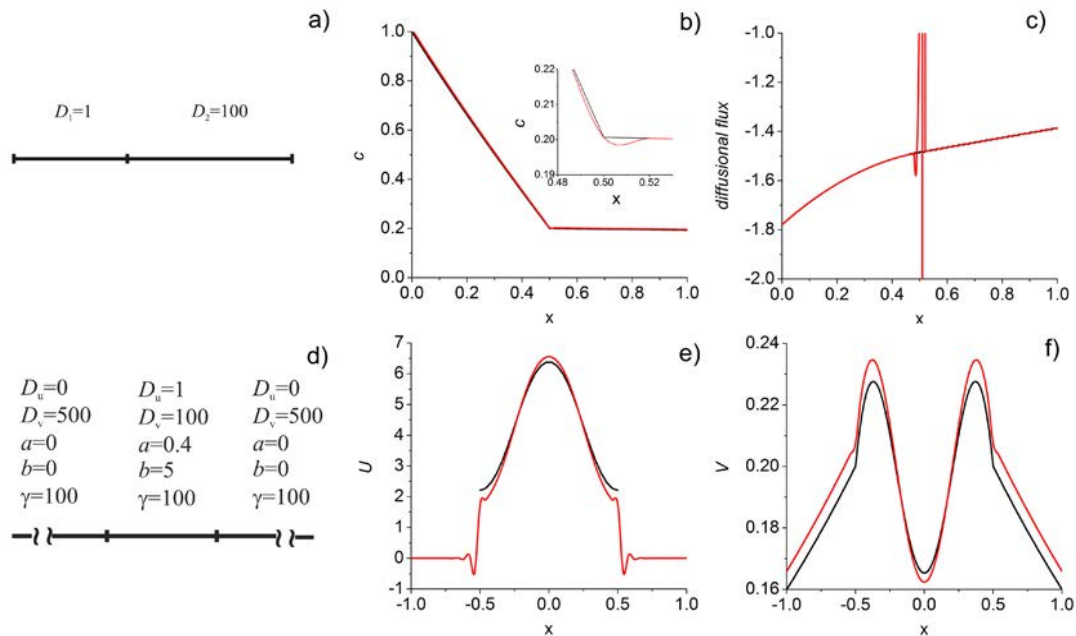

**Figure 2. Static Composite Domain.** The solutions of the simple reaction-diffusion equation 1 (upper panels) or of the Schnakenberg type equations 3 (lower panels) are more accurate if the variables are defined locally (black lines) on subdomains rather than globally (red lines) on the entire composite domains. (a,d) The computational domains on which (a) equation 1 or (d) equation 3 are solved. (b-c) The solutions of equation 1, i.e. (b) the concentration L and (c) the diffusional flux as calculated according to equation 1. (e-f) The solutions of equation 3, i.e. the concentration of (e) the slow (U) and (f) the fast (V) diffusion component respectively.

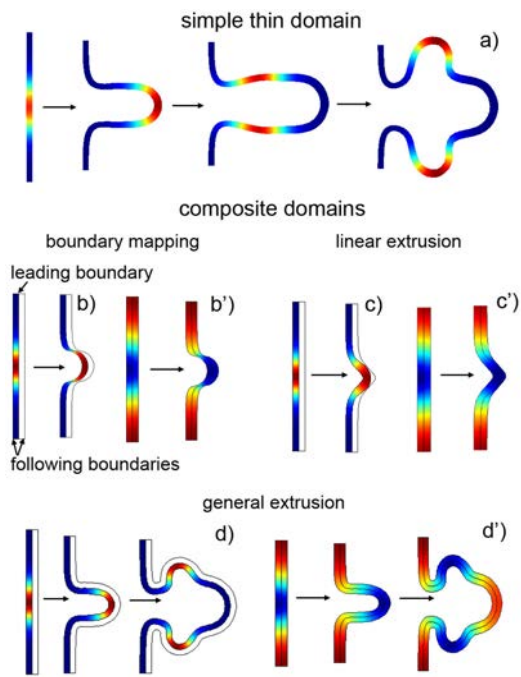

**Figure 3. Deforming growth of simple and composite domains**. (a) Deforming growth of a simple thin domain according to a concentration profile as generated by the Schnakenberg-type Turing model. (b-d) Deformations as calculated with (b) *boundary mapping*, (c) *linear extrusion* and (d) *general extrusion* for solution mapping. Panels with/without asterisk depict solution for the fast and slowly diffusing component in the Schnakenberg type equations that govern the deformations. Color code: blue and red colors correspond to low and high values respectively.

The same can be observed also for more complicated models with coupled PDEs such as the classical Turing patterning models which have been studied as models of developmental patterning processes for decades[9,10]. The classical Schnakenberg Turing model[9,11] is based on two coupled PDEs of reaction-diffusion type with variables U and V, i.e.

$$\dot{U} = \Delta U + \gamma(a - U + U^2 V)$$
$$\dot{V} = D\Delta + \gamma(b - U^2 V) \quad (3)$$

with *a, b, g, D* positive parameters and *D*>>1.

As before a global definition of the PDEs on the entire domain results in a solution with artifacts, i.e. negative concentrations (Figure 2e). The definition of equations locally on the subdomain, on the other hand, not only results in a more accurate solution, but also in a problem with a smaller number degrees of freedom. In summary, we conclude the latter approach to be superior.

### 3.2 Implicitly Defined Deformation of Composite Domains

Experimental observations show that cells and tissue can proliferate and/or move towards regions of high concentration of a morphogen.[12] We will ignore the mechanistic details of this process as these are still to be discovered and we will instead describe the local tissue movement as a function of the local concentration of a morphogen; the tissue movement is directed normal towards the surface. Figure 3a shows a solution of the Schnakenberg type equations (Equation 3) on a domain that deforms according to the rules specified above. The computational domain depicted in Figure 3a comprises only one thin domain, such that a concentration gradient can form only along the domain, but not in the perpendicular direction. If the domain is thicker or comprises several subdomains concentration gradients can develop in any direction. We would like our domain (subdomains) to deform synchronously, so that the thickness of the domain remains constant during simulation; this is often observed in nature if a deforming tissue comprises one or few monolayers of cells e.g. epithelial. To achieve synchronous deformation of subdomains even when concentration gradients can develop in any direction we need to map the solution from the leading domain border onto those, which follow (Figure 3a). COMSOL provides *Model Coupling Operators* to implement such a mapping. *Extrusion Model Coupling Operator* takes a local concentration as an argument at the following boundary and evaluates it at the corresponding point at the leading boundary. The following types of *Extrusion Coupling Operators* are available: *boundary mapping*, *linear extrusion* and *general extrusion*. As no information is available in the COMSOL documentation to help with the choice of a particular type of *extrusion coupling operator* we tested all of them. Figure 3b,c and d shows that *general extrusion* provides the desirable result – all domain boundaries deform synchronously, while *boundary mapping* and *linear extrusion* fail. In the latter case, subdomains' thickness changes over time. As a result there is a rapid degradation of mesh quality and boundaries touch each other.

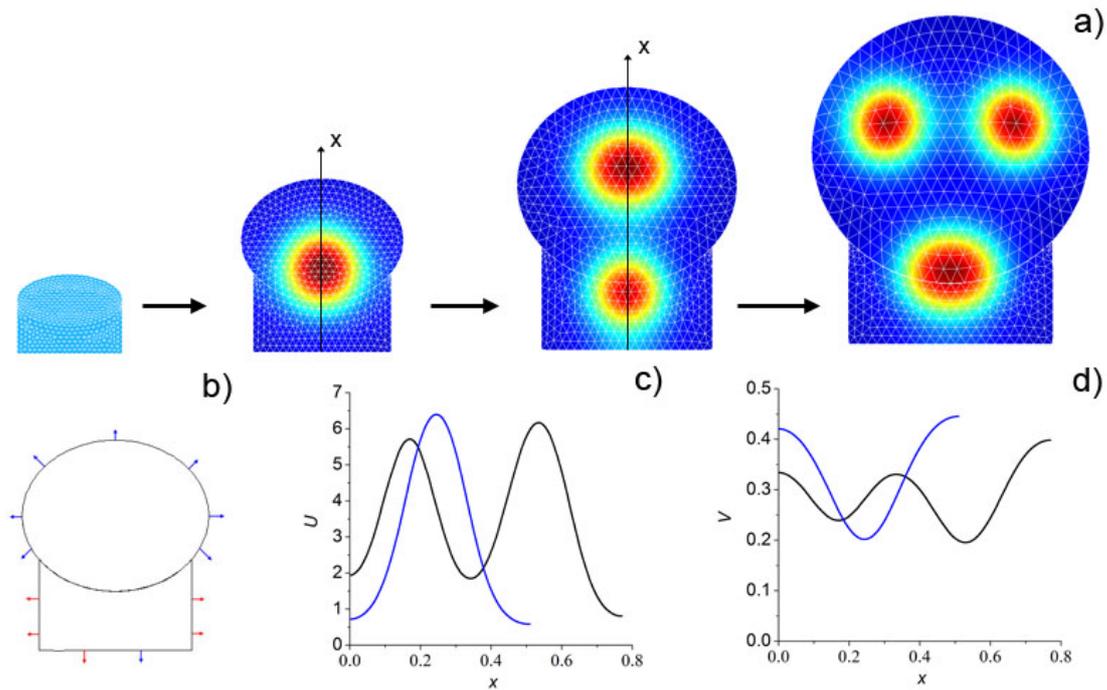

**Figure 4. Explicitly defined deformation of a composite domain.** Panel a) solution of Schakenberg type equation on a growing domain. Color code: blue and red colors correspond to low and high values, accordingly; b) shows that domain comprises ellipse and rectangular expanding with different speed; c) and d) show concentration of variable $U$ and $V$ along the lines indicated in the upper panel, blue and black lines show concentrations calculated at various times.

### 3.3 Explicitly Defined Deformation of Composite Domains

If experimental measurements are available it is desirable to solve PDEs on deforming domains where the deformation is prescribed. This is straightforward to implement using *Moving Mesh (ALE), Prescribed Mesh Displacement* unless subdomains deform according to different velocity fields (Figure 4a and b). Deformation of subdomains with uneven speed leads to the rapid degradation of mesh quality at the subdomains borders. This problem can be solved by building a computational domain not as a *union,* a default option, but as an *assembly* of subdomains. In the latter case subdomains are independent of each other. As a result meshes can be defined and deformed independently on the subdomains. To couple PDEs on the boundary of subdomains continuity of equations needs to be defined with the *Continuity* node in the *Coefficient form PDE* interface. Figure 4a shows the solution of Schnakenberg type equations (Equation 3) on the deforming composite domain comprising a rectangular and an ellipse, where the ellipse expands faster than the rectangular. Figure 4c and 4d show that the solutions for variables $U$ and $V$ are smooth.

### 4. Outlook

Here we presented a guide of how to implement models that involve deforming composite domains in COMSOL. We restricted ourselves to simplistic descriptions of tissue deformation, i.e. local growth was assumed to be normal to the surface and proportional to a local concentration of morphogen. The regulation of growth and tissue mechanics is an active field of research and the link between microscopic signaling and tissue mechanics is increasingly well explored.[13,14] As this information becomes available models will have to be developed that bridge the gap between mechanics at the micro and macro scales. This can be expected to further advance our understanding of the key mechanisms of organ development.

All our examples were defined in two spatial dimensions. Similar approaches can be used to model deforming domains in 3D (Figure 1e). However, the size of the problem and thus the computational time increases dramatically. Methods are required to enable the efficient parallelization of solvers such that computing clusters can be used efficiently. In the current COMSOL version use of multiple nodes only results in a modest speed-up.[15] Further improvements in the parallelization of the FEM solvers will be important to enable the simulation of ever more realistic and predictive models of biology.

## 5. Acknowledgements

We would like to thank the COMSOL support team and especially Sven Friedel, Rune Westin, Mikael Fredenberg, Anders Ekerot, Martin Nilsson for their support and insightful comments, Rolf Zeller and his group (and in particular Erkan Ünal) for the experimental data and the ongoing collaboration, and Amarendra Badugu for experimental data processing. The authors acknowledge funding from an ETH Zurich postdoctoral fellowship to D.M. and funding from the SNF Sinergia grant "Developmental engineering of endochondral ossification from mesenchymal stem cells".